\documentclass{article}
\usepackage[margin=1.0in]{geometry}

\usepackage[sorting=none,citestyle=phys]{biblatex}
\usepackage{pdfpages}

\usepackage{graphicx}
\usepackage{bm}
\usepackage{color}
\usepackage{xcolor}
\usepackage{amsmath}
\usepackage{amstext}
\usepackage{amssymb}
\usepackage{booktabs}
\usepackage{array}
\usepackage{ulem}
\usepackage{hyperref}
\usepackage{verbatim} 

\hypersetup{
colorlinks,
citecolor=blue,
filecolor=blue, 
linkcolor=blue, 
urlcolor=blue
}

\renewcommand{\emph}[1]{\textit{#1}}

\newcommand{\editor}[2]{%
  \expandafter\newcommand\csname #1note\endcsname[1]{%
    \textcolor{#2}{(\textbf{#1:} ##1)}}%
  \expandafter\newcommand\csname #1\endcsname[1]{%
    \textcolor{#2}{##1}}%
  \expandafter\newcommand\csname #1cancel\endcsname[1]{%
    \textcolor{#2}{\sout{##1}}}%
  \expandafter\newcommand\csname #1change\endcsname[2]{%
    \textcolor{#2}{\sout{##1}##2}}%
  \newenvironment{#1text}{\color{#2}}{\color{black}}
}
\definecolor{Blu}{rgb}{0.00,0.00,1.00}
\definecolor{Red}{rgb}{1.00,0.00,0.00}
\definecolor{Cyan}{rgb}{0.00,0.50,0.50}
\definecolor{Green}{rgb}{0.00,0.70,0.00}
\editor{AG}{orange}
\editor{DAL}{Green}
\editor{FQ}{Cyan}

\title{Estimating the number of available states 
for normal and tumor tissues in gene expression space}
\author{Augusto Gonzalez$^{1,2}$, Frank Quintela$^{3,2}$, Dario A. Leon$^{4,2}$, Maria Luisa Bringas Vega$^{1,5}$, \\ Pedro Valdes Sosa$^{1,5}$}
\date{
$^1$University of Electronic Science and Technology, 610051, Chengdu (People Republic of China) \\
$^2$Institute of Cybernetics, Mathematics and Physics, 10400, Havana (Cuba) \\
$^3$University of Modena \& Reggio Emilia, 41125, Modena, (Italy)  \\
$^4$S3 Centre, Istituto Nanoscienze, CNR, 41125, Modena (Italy) \\
$^5$Cuban Neurosciences Center, 11600, Havana, (Cuba) \\
\today
}

\begin{document}
\maketitle

\begin{abstract}
The topology of gene expression space for a set of 12 cancer types is studied by means of an entropy-like magnitude, which allows the characterization of the regions occupied by tumor and normal samples. The comparison indicates that the number of available states in gene expression space is much greater for tumors than for normal tissues, suggesting the irreversibility of the progression to the tumor phase. The entropy is nearly constant for tumors, whereas exhibits a higher variability in normal tissues, probably due to tissue differentiation.
In addition, we show an interesting correlation between the fraction of available states and the overlapping between the tumor and normal sample clouds, interpreted as a way of reducing the decay rate to the tumor phase in more ordered or structured tissues. 
\end{abstract}

\section{Introduction}

The extreme difficulties in treating cancer~\cite{cancer} reveal that the survival capabilities of cancer cells are
much stronger than those of the somatic cells in our body, restricted by the conditions of homeostasis. 
The reason for such ``advantages'' is explained in the atavistic theory of cancer~\cite{atavistic1,atavistic2,atavistic3,atavistic4,atavistic5} as the result of a core
genetic programme, which helped primitive multicellular organisms to overcome the extreme conditions posed by the ancient earth.  

One aspect of these enhanced capabilities is related to tissue fitness. Cancer cells are known to turn
off the mechanism of fitness control in homeostasis and exhibit higher replication rates than stem cells in healthy tissues \cite{replicrates}.

In vivo measurements of fitness in normal and tumor tissues could be a difficult task. However, 
there is a way of looking at fitness which is related to the number of available states 
for a system in phase space and may be the subject of numerical computations. 
Indeed, for a tissue (or a small portion of it) there should be a fitness landscape 
in gene expression space~\cite{detbal}. Regions of high fitness are characterized by their volumes which should be proportional to the number of available states for the system. 

In the present paper, we aim at estimating the number of available states for normal and tumor tissues or, 
more precisely, the ratio of numbers for the tumor and the corresponding normal tissue. To this end, 
we process gene expression data for 12 types of cancer, coming from The Cancer Genome Atlas (TCGA) 
portal~\cite{tcga}. 
Notice that, in the TCGA data, gene expression levels are measured in small tissue samples, obtained from biopsies. Although modern techniques 
allow measuring the expressions in individual cells~\cite{gecells}, we stress that a micro-sample contains the information coming from many different cells 
and their interactions. With the purpose of estimating the relative fitness, 
the comparison between pathologically cancerous and normal samples is 
meaningful and realistic. 

The idea to measure the number of available states is to use an analogy with Statistical Physics~\cite{ergodic} or Semiclassical Mechanics~\cite{Semiclass} in which this number is proportional to the volume spanned by the system in phase space. In our biology problem, we understand that GE space is a kind of configuration space, and the fitness landscape plays the role of external potential in Physics~\cite{detbal}.

The normal and tumor regions in GE space define attractors~\cite{detbal}, that is local maxima of fitness. They are separated by a low fitness barrier. As it will become apparent below, normal and tumor samples from the TCGA data are distributed around their respective attractors. These samples come from different individuals, each one with a particular history of tumor progression. By using an analogy with the ergodic principle~\cite{ergodic}, we assume that the actual distribution of samples is a phase portrait of the trajectories of an ensemble of microstates that start in the normal region. Some of these microstates progress to the tumor zone. Thus, we fit the observed distributions with gaussians (the functions with lowest bias from the point of view of information theory), and ``compute'' the volumes (hipervolumes) of their basins of attraction by means of an entropy-like magnitude that is roughly the logarithm of the volume. 

The computation has subtle details, in particular the dimensionality. By using data on the variance distribution and a criterium from information theory, we manage to truncate the Principal Component expansion \cite{pca1,Joan} to the first 20 components for the 12 studied tumors, which allows the comparison not only between the volumes of the normal and tumor basins of attraction, but also among different tissues. 

In addition, the computed density distributions allow to estimate the overlap between normal and tumor clouds of samples. This magnitude shows an interesting correlation with the ratio of basin volumes.

Let us stress that we have estimated the configuration space volumes. The ``dynamical'' component (i.e. momentum space analog) is still missing. As mentioned above, we expect higher replication rates for tumors, that is higher momentum space volumes. However, at present this information is lacking and we can not present reliable numbers.

The main results of our paper are the following. First, the number of available states is much higher for tumors than for
normal tissues. This may be expected since an homeostatic tissue has much less possibilities of realization or a more constrained order 
than the primitive multicellular tumor. Second, the entropy of tumors takes a nearly constant value, a fact consistent with their common evolutionary origin in the atavistic theory. Normal tissues, on the contrary, exhibit a higher
variability of their entropy, probably a manifestation of tissue differentiation. And third, 
there is a correlation between the ratio of basin volumes and the overlapping between the normal and tumor clouds of samples, indicating a non trivial topology of gene expression space
aimed at reducing the decay rate (the cancer risk) of more ordered or structured tissues. These facts are discussed below.

\section{Results and Discussion}

\subsection{Entropy in gene expression space}
As mentioned, our starting point is the TCGA expression data for 12 tumors and the corresponding normal tissues. The selected types of cancer are characterized by more than 20 normal and more than 300 tumor samples, as shown in Table \ref{tab1}.

\begin{table}
  \begin{center}
    \begin{tabular}{|l|l|l|l|l|l|l|} 
      \hline
      Tissue & Normal samples & Tumor samples & $\Delta S$ & $S_{tumor}$ & $- \ln I$ \\
      \hline
      \hline
      BRCA & 112 & 1096 & 17.93$\pm$1.12 & 89.16 & 13.25$\pm$1.51 \\
      \hline
      COAD & 41  & 473  & 29.93$\pm$2.33 & 90.87 & 28.15$\pm$6.19  \\
      \hline
      HNSC & 44  & 502  & 12.67$\pm$0.99 & 89.48 & 10.32$\pm$1.05  \\
      \hline
      KIRC & 72  & 539  & 22.23$\pm$1.53 & 89.31 & 17.51$\pm$1.43  \\
      \hline
      KIRP & 32  & 289  & 29.93$\pm$1.97 & 90.40 & 25.89$\pm$6.16  \\
      \hline
      LIHC & 50  & 374  & 29.00$\pm$0.93 & 90.16 & 16.12$\pm$3.94  \\
      \hline
      LUAD & 59  & 535  & 26.24$\pm$1.18 & 92.18 & 16.21$\pm$2.40  \\
      \hline
      LUSC & 49  & 502  & 28.59$\pm$0.99 & 90.58 & 21.84$\pm$4.95  \\
      \hline
      PRAD & 52  & 499  & 9.79$\pm$1.41 & 85.96 & 6.60$\pm$0.67  \\
      \hline
      STAD & 32  & 375  & 19.82$\pm$1.44 & 93.10 & 14.12$\pm$4.04  \\
      \hline
      THCA & 58  & 510  & 15.00$\pm$0.87 & 85.62 & 9.77$\pm$1.59  \\
      \hline
      UCEC & 23  & 552  & 26.06$\pm$2.43 & 93.26 & 22.06$\pm$5.14  \\
      \hline
    \end{tabular}
  \end{center}
\caption{The set of studied cancer types and the main 
results of the paper. 
TCGA abbreviations and details of how error bars are estimated are explained in the Supplementary Information.}
    \label{tab1}
\end{table}

We perform a Principal Component Analysis (PCA) \cite{pca1,pca2,pca3} of the expression data. Methodological aspects are detailed in paper \cite{AYR}, where we study the topology of gene expression space for normal and tumor tissues. For completeness, we sketch the main results of that paper that shall be used in our computations. Details can be found in the Methods section and the Supplementary Information, in particular:

1. Although there are around 60000 genes, normal tissues and tumors span a region with reduced effective dimension. 
Then, we use the first 20 principal components in order to describe the state of a sample in gene expression space (GES). These 20 components capture no less than 85 \% of the total variance in the dispersion of experimental samples in GES, and practically saturate the Akaike Information Criterium~\cite{Akaike}.

2. For a given tissue, normal samples are well separated from tumor samples in GES. Both regions seem to be the basins of attraction of two singular points: the normal homeostatic and the cancer attractors.

\begin{figure}
\begin{center}
\includegraphics[width=0.45 \linewidth]{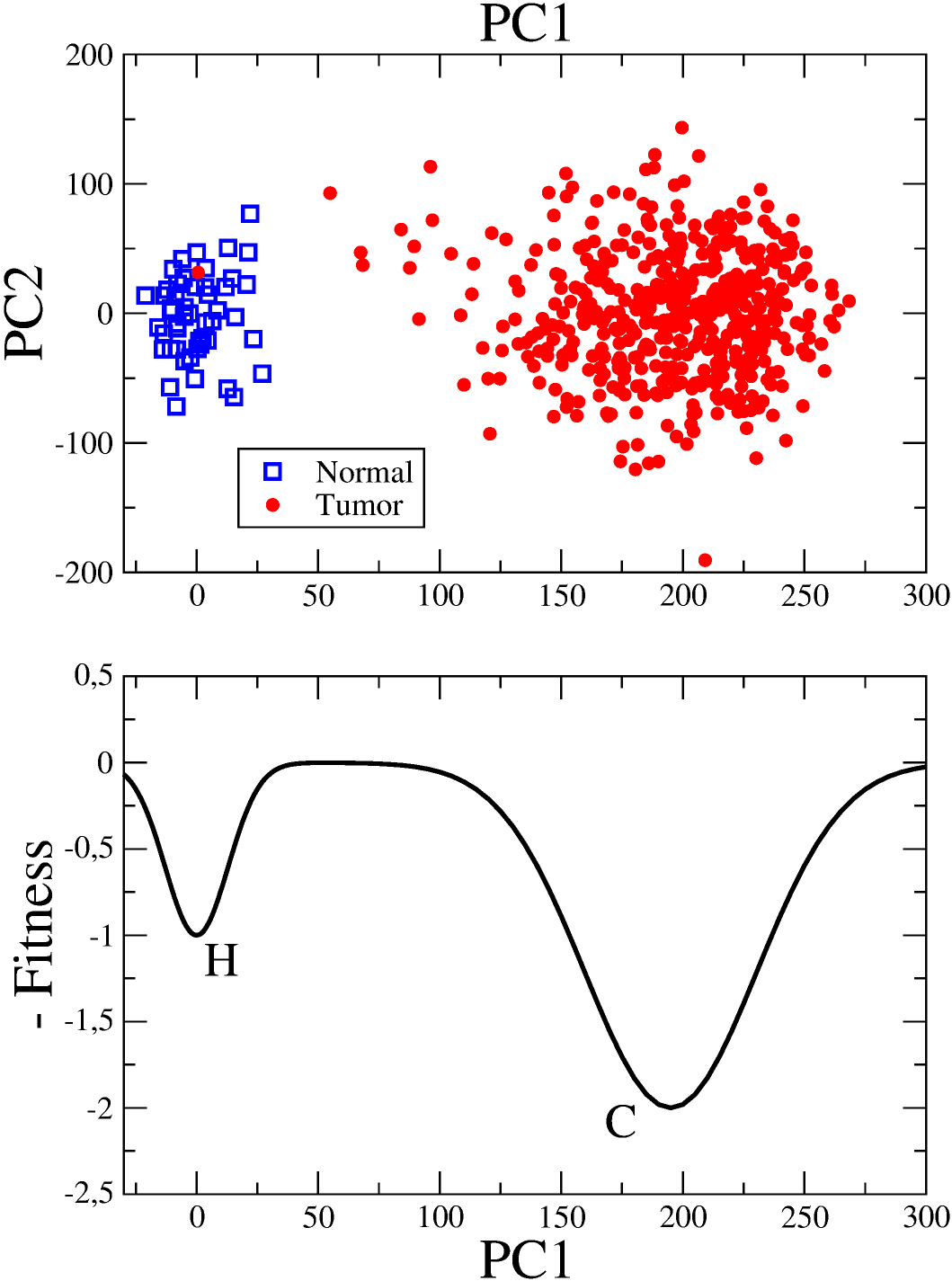}
\end{center}
\caption{Upper panel: PCA of gene expression data for Squamous Cell Lung cancer (LUSC). The position along the first axis (PC1) discriminates between a normal sample and a tumor. Lower panel: Schematics of the fitness landscape. The $x$ axis is again PC1, but the $y$ axis represents the fitness with a minus sign. H and C labels the normal (homeostatic) and cancer states, respectively. The maximum fitness in the H state is normalized to unity.}
\label{fig1} 
\end{figure}

Fig. \ref{fig1} upper panel shows as example the (PC1, PC2) plane for Lung Squamous Cell cancer (LUSC in TCGA notations). Points in the figure represent samples from different patients. The clouds of
points are grouped in well defined regions defining the attractors. We shall estimate the volume of each region, which gives an indication of the number of accessible states.

More precisely, for both normal tissues and tumors we shall introduce the entropy-like magnitude \cite{entropy}:

\begin{equation}
S=-\int d^{D}\underline{x}~\rho(\underline{x})\ln \rho(\underline{x}),
\label{eq1}
\end{equation}

\noindent
where $D=20$ is the number of principal components to be used in the description of the system in GES, and $\rho$ is the probability density, normalized to unity, coming from a fit to the observed sample data. 

The relation between the $S$ magnitude and the volume, $V$, of the basin of attraction is roughly $S\approx \ln (V)$, thus $S$ measures the logarithm of the number of available states in the region.

We fit the observed distribution of sample points to a multivariate gaussian density, $\rho$. This procedure guarantees a maximal entropy, as compared to any other possible ansatz for $\rho$, and a minimal bias from the point of view of information theory.

We show in Table \ref{tab1} the magnitudes $S_{tumor}$ and $\Delta S=S_{tumor}-S_{normal}$ for the set of tissues under study. The way that bar errors are estimated is described in the Supplementary Information.

The number of states in GES seems to be much larger for tumors
than for normal tissues, leading to $\Delta S >> 1$.

On the other hand, the number of accessible states appears to be nearly constant for all tumors. Normal tissues exhibit larger variations, which could be perhaps related to tissue differentiation. In other words, the process of de-differentiation of tumors~\cite{dediff} seems to involve the
increase of the accessible volume in GES to a nearly constant value.

We can not afford a rationale for the computed entropies of normal tissues, that is lower values in epithelial tissues for example, nor relate the entropies to their developmental origin. We can neither relate the normal state entropy to the risk of cancer in the tissue. The prostate (PRAD), a tissue in which cancer is very common,  seems to exhibit the higher disorder (entropy), but lung (LUAD, LUSC) and colon (COAD), with much lower entropies,  are also high risk tissues. Naively, one would expect the entropy difference, not the entropy in the normal state, to correlate with the cancer risk. Indeed, more available tumor states should indicate more probability to transit to the tumor region. The question is, however, much more subtle, as shown in the next sections.

\subsection{Cloud overlapping}
We may introduce an additional magnitude characterizing the transition region between the two attractors, that is the overlapping between the clouds of normal and tumor samples. 

Let us define the density overlap:

\begin{equation}
I=\int d^{D}\underline{x}\sqrt{\rho_{tumor}(\underline{x}) \rho_{normal}(\underline{x})}.
\label{eq2}
\end{equation}  

\noindent
The square root is introduced for normalization purposes. The analytic expression for $I$, when the $\rho$ are gaussian distributions, is provided in the Methods section below. 

\begin{figure}
\begin{center}
\includegraphics[width=0.96 \linewidth]{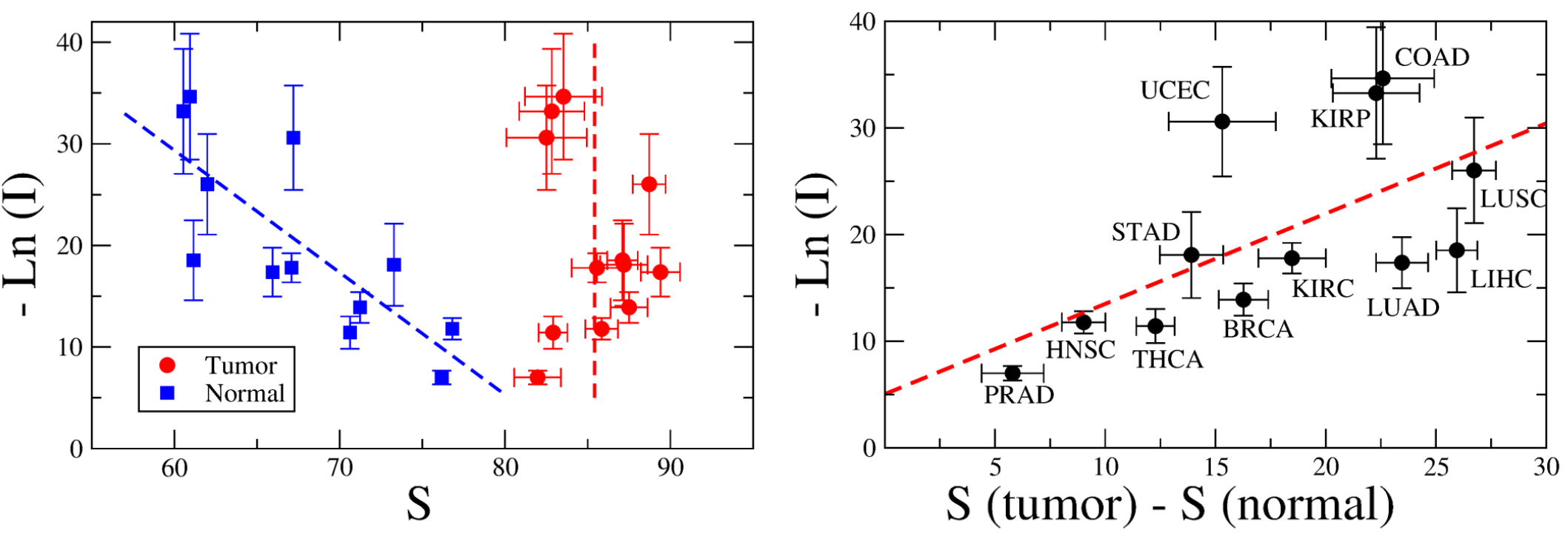}
\end{center}
\caption{The entropy-overlapping map. Notice that tumors exhibit a nearly constant entropy, and that
there is an exponential relationship between the overlap $I$ and the entropy variation $\Delta S$.}
\label{fig2}
\end{figure}

The results of computations are shown in Table~\ref{tab1} and Fig.~\ref{fig2} for the set of 12 tissues studied in the present paper. 
Fig.~\ref{fig2}, in which we plot cloud overlapping vs entropy, can be understood as a complexity map~\cite{complexmap} for different normal tissue - tumor pairs.

The observed overlap could intuitively be related to the distance between the centers of the clouds. The distances along the first PC axis, PC1, are computed in paper~\cite{Joan}. These computations confirm, for example, that in PRAD the cloud centers are much closer than in COAD or LUSC.

The numbers in Table~\ref{tab1} and Fig.~\ref{fig2} indicate also the apparent correlation between $-\ln I$ and $\Delta S$, i.e. $-\ln I = 0.85~ \Delta S - 1.95$ or $I\propto (V_n/V_t)^{0.85}$. The p-value of the linear fit in the log-log plot for these magnitudes is 0.02.

The nature of this dependence is intriguing. The fact is that 
the larger the entropy difference (the ratio of basin volumes) the smaller the overlapping between the tumor and normal sample clouds. An interpretation for this fact is provided in the next section.

\subsection{Fitness landscape and transition rates}

The normal homeostatic state shall be protected against transitions to the cancer state by a barrier. Otherwise the transitions are unavoidable because both the fitness and the number 
of available states in the cancer region are much higher than in the normal region.

It is natural to assume that the intermediate region holds a low-fitness barrier, as schematically represented in Fig.~\ref{fig1} lower panel for Lung
Squamous Cell Cancer (LUSC). Indeed, the normal homeostatic state is a state with regulated fitness~\cite{regfitness}. In cancer, on the other hand, these constraints are removed and tumor growth is only limited by the availability of space and nutrients. The intermediate region is a space for senescence or different kinds of illness, where fitness is reduced and the compensation
mechanisms are not capable of keeping homeostasis. 


In Fig.~\ref{fig1} lower panel we provide a schematic 1D representation of the fitness landscape. The x axis, as in the upper panel, is PC1, which is identified as the cancer axis~\cite{AYR}. The normal and cancer states are well separated along this axis. The y axis, on the other hand, is a sketch for the fitness (with a minus sign), which is obtained simply by smoothing the histogram of samples. In other words, we assume that the observed density of samples at a given point of GES is proportional to the fitness.

The absolute maximum of fitness is at the cancer attractor (denoted C in the figure). The normal homeostatic state (H) is a local metastable maximum, which should be characterized by a mean decay time, $\tau_H$. In the figure, the fitness at the homeostatic maximum is normalized to unity.
Notice that with a rough estimation of the fitness landscape 
we could get, in principle, an estimation for $\tau_H$, and thus the risk of cancer in a tissue. 

The time for the reverse process to occur, $\tau_C$, that is from the tumor to the normal state, is expected to be much larger than $\tau_H$. We could get a rough value for it  by using a kind of detailed balance equation \cite{detbal}: 

\begin{equation}
\tau_C = \tau_H \frac{N_{states}(C)}{N_{states}(H)} = \tau_H \exp (\Delta S).
\label{eq3}
\end{equation}

Eq.~\eqref{eq3} forces the product of  the decay rate ($1/\tau$) and the number of micro-states to be equal in both states, H and C.
Taking $\tau_H\approx 60$ years we get for prostate tumors, for example, $\tau_c\approx 10^6  \text{ years} \equiv 1 \text{ My}$.
For thyroid cancer, on the other hand, $\tau_C\approx 200 \text{ My}$. 
These are fictitious numbers, not related to any biological processes. We compute them with the
only purpose of confirming that the progression to cancer is an almost irreversible process. 

On the other hand, it is a curious fact that the required times for early multicellular organisms to evolve to modern metazoans are precisely hundreds of My~\cite{times}. At the level of conglomerates of cells one can
imagine evolution as jumps against entropy, that is, from states like C to states like H. These are highly improbable processes which, however, may be the source of further advantages at a different level of organization. When one says that it may take 200 My to occur, it means that from the many cell  conglomerates living in this time period a few of them could make the transition and start a new line of evolution.

Eq.~\eqref{eq3} for the decay time lacks of an important factor: the effect of 
the barrier, which is related to the magnitude $I$. Wider barriers, corresponding to lower values of $I$, that is higher $-\ln I$, should slow
down the transitions. From this perspective, the correlations between $-\ln I$
and $\Delta S$ is quite natural. Let us use the same Eq.~\eqref{eq3}, but now taking $\tau_C\approx 200 \text{ My}$ as a reference in order to estimate $\tau_H$. A more ordered tissue (greater entropy difference) has a smaller $\tau_H$, and should be separated from the tumor by a wider barrier in order to prevent the transitions. This argument, although qualitative and preliminary, 
indicates a possible very interesting relation between the topology of gene expression space (volumes and intersections) and the decay rate, which is related to the risk of cancer in a tissue.

\subsection{Concluding remarks}

We initiated in papers \cite{AYR,Joan} a quantitative study of the topology of GES in tumors. In particular, the distances between the center of the tumor and normal regions and their r.m.s. radii along the PC1 axis were computed, and were shown to correlate with the characteristics of the GE distribution functions~\cite{Joan}.  

In the present paper we deal with two more magnitudes quantifying the topology of GES. First, we estimated the volumes (hypervolumes) of the basins of attraction for the normal and cancer regions in each of the 12 types of cancer described in Table~\ref{tab1}. There are subtle details concerning the computation of these quantities which are discussed in the Supplementary Information. The most important one is the effective dimension of the regions. We have used the variance distribution in the PC analysis and ideas from information theory~\cite{Akaike} in order to define the effective dimension, and thus to compute the volumes. We use the same effective dimension, 20, for all of the tissues in such a way that they may be compared. 

Using an analogy from Statistical Physics~\cite{ergodic} and semiclassical Quantum Mechanics~\cite{Semiclass}, in which the volume of phase space is related to the number of states, we have related the computed volumes to the number of accessible biological micro-states. Volumes are measured by means of a ``configurational'' entropy-like magnitude, constructed from the probability density of samples in the space. The latter is obtained from a multivariate gaussian fit to the observed distribution of samples.

The second magnitude characterizing the topology of GES is the overlapping between the normal and tumor clouds, computed from the same probability densities.

The results of the paper are mainly three: 1. The number of accessible states is much higher for tumors than for normal samples, 2. All studied tumor localizations have roughly the same number of accessible states whereas normal tissues exhibit higher variability, and 3. The overlap between the tumor and normal samples clouds of points is roughly proportional to $\exp (-0.85~\Delta S)$. 

The reduced number of accessible states for the normal tissue can be interpreted as a higher level of organization, as compared to the tumor. The nearly constant entropy of tumors points to the common evolutionary origin of tissues, in accordance with the atavistic theory. The higher variability of entropy in normal tissues, on the other hand, can be taken as a manifestation of tissue differentiation and structure. Finally, the correlation between cloud overlapping and the entropy difference is interpreted as a way of slowing down the transition to the cancer state in more organized tissues, indicating a possible very interesting relation between the topology of gene expression space (volumes and intersections) and the risk of cancer in a tissue.

The results seem consistent with the fundamentals of evolution theory and the atavistic theory of cancer.

\section{Methods}
\subsection{Principal component analysis}
The TCGA data for the tissues described in Table 1 is analyzed by means of the PCA technique. The details of the PC analysis are described 
in paper \cite{AYR}. We briefly sketch them in the present section.   

Gene expression are given in FPKM format. The number of genes is 60483. This is the dimension of matrices in the PCA processing.

We take the mean geometric average over normal samples in order to define the reference expression for each gene, $e_{ref}$. Then the normalized or differential expression is defined as: $e_{diff} = e/e_{ref}$. The fold variation is defined in terms of the logarithm $\hat e = \log _2 (e_{diff})$. Besides reducing the variance, the logarithm allows treating over- and sub-expression in a symmetrical way. 

Deviations and variances are measured with respect to $\hat e = 0$. That is, with respect to the average over normal samples. This election is quite natural, because normal samples are the majority in a population. 

With these assumptions, the covariance matrix is written:

\begin{equation}
\sigma_{ij} = \frac{1}{N_{samples}-1} \sum \hat e_i(s) \hat e_j(s), 
\label{eq6}
\end{equation}

\noindent
where the sum runs over the samples, $s$, and $N_{samples}$ is the total number of samples (normal plus tumor). $\hat e_i(s)$ is the fold variation of gene $i$ in sample $s$.

As mentioned, the dimension of matrix $\sigma$ is 60483. By diagonalizing it, we get the axes of maximal variance: the Principal Components (PCs). They are sorted in descending order of their contribution to the variance. 

In LUSC, for example, PC1 accounts for 67\% of the variance. This large number is partly due to our choice of the reference,  $\hat e = 0$, and the fact that most of the samples are tumors. The reward is that PC1 may be defined as the cancer axis.  The projection over PC1 defines whether a sample is classified as normal or tumor. 

The next PCs account for a smaller fraction of the variance. PC2 is responsible of 4\%, PC3 of 3\%, etc.  Around 20 PCs are enough for an approximate description of the region of the gene expression space occupied by the set of samples.

\subsection{Entropy and overlapping integral
\label{sec:EandOI}}

For a sample, the projections over the PC vectors define the new coordinates. 
These are the starting data for the computation of the configurational entropy. We organize it as $24$  matrices $M$, each one corresponding to a tissue in a stage, for example $M(\text{LIHC, tumor})$. The number of columns in any case is 20 (number of Principal Components) and the number of rows is the number of samples, as reported
in Table ~\ref{tab1}. 

From $M$ the sample covariance matrix, $\Sigma$, is defined as 

\begin{equation}
 \Sigma_{jk} = \frac{1}{N-1}\sum^{N}_{i=1}(M_{ij}-\mu_{j})(M_{ik}-\mu_{k}),
\label{eq7}
\end{equation} 

\noindent
where $\mu_j=\frac{1}{N}\sum_{i=1}M_{ij}$ is the mean value of coordinate $j$ in the set of samples.

In order to find probability distributions for the sets of normal and tumor samples we maximize the entropy taking the covariance
matrices as constraints. These are quadratic constraints, thus the result is a multivariate gaussian \cite{gaussian}:

\begin{equation}
 \rho(\underline{x}) = \frac{1}{(2\pi)^{\frac{D}{2}}\sqrt{|\Sigma|}}\exp \left[-\frac{1}{2}(\underline{x}-\underline{\mu})^{T} \Sigma^{-1}(\underline{x}-\underline{\mu})\right].
 \label{eq:rho}
\end{equation}

Notice our convention for vectors, $\underline{x}$. There are advantages in using this procedure.  First, with normal distributions we may analytically compute the quantities of interest, second this distribution is, in accordance with the Central Limit Theorem, 
an estimation of the actual distribution for much larger data sets, and third this distribution is, from the point of view of information theory, the most unbiased 
one with regard to data covariance, that is no heuristic criteria have been used for choosing it. 

For our target quantities, the entropy and the overlap integral, we get:

\begin{equation}
 \label{eq:entropy}
 S=\frac{1}{2}\ln |\Sigma| + \frac{D}{2}\left(1+\ln 2\pi\right),
\end{equation}

\begin{equation}
 \begin{split}
 I=2^{\frac{D}{2}}\frac{|\Lambda_n|^{1/4}|\Lambda_t|^{1/4}}{|\Lambda_c|^{1/2}} 
 \exp \Bigg[\frac{1}{4} ( & \eta_{c}^T\Lambda_c^{-1}\eta_{c} -  \mu_n^T\Lambda_n\mu_n-\mu_t^T\Lambda_t\mu_t) \Bigg],
 \end{split}
 \label{eq:overlap}
\end{equation}

\noindent 
where $\Lambda_{j}=\Sigma^{-1}_{j}$ for $j=n,t$; 
$\eta_{c} = \Lambda_{n} \mu_{n}+\Lambda_{t} \mu_{t}$, and  $\Lambda_{c}=\Lambda_{n}+\Lambda_{t}$.

Details on the dependence of $S$ and $I$ on the number of samples used in their computation are provided in the Supplementary Information. 

\printbibliography

@ARTICLE{cancer,
   author = {Victoria da~Silva-Diz and Laura Lorenzo-Sanz and Adrià Bernat-Peguera},
   title = {Cancer cell plasticity: Impact on tumor progression and therapy response},
   journal = {Semin. Cancer Biol.},
   volume = {53},
   pages = {48-58},
   year = {2018},
   doi = {https://doi.org/10.1016/j.semcancer.2018.08.009}
}

@ARTICLE{atavistic1,
   author = {Davies, Paul~C.~W. and Lineweaver, Charles~H.},
   title = {Cancer tumors as Metazoa 1.0: tapping genes of ancient ancestors},
   journal = {Phys. Biol.},
   volume = {8 (1)},
   pages = {015001},
   year = {2011},
   doi = {https://doi.org/10.1088/1478-3975/8/1/015001}
}

@ARTICLE{atavistic2,
   author = {Tomislav Domazet-Lošo and Diethard Tautz},
   title = {Phylostratigraphic tracking of cancer genes suggests a link to the emergence of multicellularity in metazoa},
   journal = {BMC Biology},
   volume = {8 (1)},
   pages = {66},
   year = {2010},
   doi = {https://doi.org/10.1186/1741-7007-8-66}
}

@ARTICLE{atavistic3,
   author = {Charles~H. Lineweaver and Paul~C.~W. Davies and Mark~D. Vincent},
   title = {Targeting cancer’s weaknesses (not its strengths): Therapeutic strategies suggested by the atavistic model},
   journal = {Bioessays},
   volume = {36 (9)},
   pages = {827-835},
   year = {2014},
   doi = {https://doi.org/10.1002/bies.201400070}
}

@ARTICLE{atavistic4,
   author = {Cisneros, Luis and Bussey, Kimberly~J. and Orr, Adam~J. and Miocevic, Milica and Lineweaver, Charles~H. and Davies, Paul},
   title = {Ancient genes establish stress-induced mutation as a hallmark of cancer},
   journal = {Front. Cell. Dev. Biol.},
   volume = {12 (4)},
   pages = {e0176258},
   year = {2017},
   doi = {https://doi.org/10.1371/journal.pone.0176258}
}

@ARTICLE{atavistic5,
   author = {Anna~S. Trigos and Richard~B. Pearson and Anthony~T. Papenfuss and David~L. Goode},
   title = {Somatic mutations in early metazoan genes disrupt regulatory links between unicellular and multicellular genes in cancer},
   journal = {ELife},
   volume = {8},
   pages = {e40947},
   year = {2019},
   doi = {https://doi.org/10.7554/eLife.40947.001}
}

@BOOK{replicrates,
   author = {Bruce Alberts and Dennis Bray and Karen Hopkin and Alexander Johnson and Julian Lewis and Martin Raf and Keith Roberts and Peter Walter},
   title = {Essential cell biology},
   publisher = {Garland Science},
   isbn = {978-0-8153-4454-4},
   year = {2013}
}

@ARTICLE{tcga,
  author = { Katarzyna, Tomczak and Patrycja, Czerwińska and Maciej, Wiznerowicz},
  title = {The {C}ancer {G}enome {A}tlas ({TCGA}): an immeasurable source of knowledge},
  journal = {Contemp Oncol (Pozn)},
  volume = {19(1A)},
  pages = {A68–A77},
  year = {2015},
  doi = {https://doi.org/10.5114/wo.2014.47136}
}

@ARTICLE{gecells,
   author = {Paul~C. Blainey and Stephen~R. Quake},
   title = {Dissecting genomic diversity, one cell at a time},
   journal = {Nat. Methods},
   volume = {11},
   pages = {19-21},
   year = {2014},
   doi = {https://doi.org/10.1038/nmeth.2783}
}

@Article{pca1,
  author   = {Svante Wold and Kim Esbensen and Paul Geladi},
  journal  = {Chemometrics and Intelligent Laboratory Systems},
  title    = {Principal component analysis},
  year     = {1987},
  issn     = {0169-7439},
  number   = {1},
  pages    = {37-52},
  volume   = {2},
  doi      = {https://doi.org/10.1016/0169-7439(87)80084-9}
}

@Article{pca2,
  author  = {Lever, Jake and Krzywinski, Martin and Altman, Naomi},
  journal = {Nature Methods},
  title   = {Principal component analysis},
  year    = {2017},
  month   = jun,
  pages   = {641-642},
  number  = {7},
  volume  = {14},
  doi     = {https://doi.org/10.1038/nmeth.4346}
}

@Article{pca3,
  author  = {Ringnér, Markus},
  journal = {Nature biotechnology},
  title   = {What is principal component analysis?},
  year    = {2008},
  month   = {04},
  pages   = {303-304},
  volume  = {26},
  doi     = {https://doi.org/10.1038/nbt0308-303}
}

@MISC{AYR,
   author = {Gonzalez, Augusto and Perera, Yasser and Perez, Rolando},
   title = {On the gene expression landscape of cancer},
   howpublished = {\url{https://arxiv.org/abs/2003.07828v3}},
   year = {2020}
}

@BOOK{entropy,
   author = {Thomas~M. Cover and Joy~A. Thomas},
   title = {Elements of Information Theory, 2nd Edition},
   publisher = {Wiley-Interscience},
   isbn = {978-0-471-24195-9},
   howpublished = {\url{https://www.wiley.com/en-it/Elements+of+Information+Theory,+2nd+Edition-p-9780471241959}},
   year = {2006}
}

@Article{dediff,
  author  = {Dinorah Friedmann-Morvinski and Inder~M. Verma},
  title   = {Dedifferentiation and reprogramming: origins of cancer stem cells},
  journal = {EMBO reports},
  volume  = {15 (3)},
  pages   = {244-253},
  year    = {2014},
  doi     = {https://doi.org/10.1002/embr.201338254}
}

@Article{complexmap,
  author  = {David~P. Feldman and Carl~S. McTague and James~P. Crutchfield},
  title   = {The organization of intrinsic computation: Complexity-entropy diagrams and the diversity of natural information processing},
  journal = {Chaos},
  volume  = {18 (4)},
  pages   = {043106},
  year    = {2008},
  doi     = {https://doi.org/10.1063/1.2991106}
}

@Article{Joan,
   author = {Gonzalez, Augusto and Nieves, Joan and Leon, Dario~A. and Bringas, Maria~Luisa and Valdes-Sosa, Pedro},
   title = {Gene expression rearrangements denoting changes in the biological state},
  journal = {Sci Rep},
  volume  = {11},
  pages   = {8470},
  doi     = {https://doi.org/10.1038/s41598-021-87764-0},
  year = {2021}
}

@Article{regfitness,
  author  = {Benoit Biteau and Christine~E. Hochmuth and Heinrich Jasper},
  title   = {Maintaining Tissue Homeostasis: Dynamic Control of Somatic Stem Cell Activity},
  journal = {Cell Stem Cell},
  volume  = {9},
  pages   = {402-411},
  year    = {2011},
  doi     = {https://doi.org/10.1016/j.stem.2011.10.004}
}

@Article{detbal,
  author  = {Jin Wang},
  title   = {Landscape and flux theory of non-equilibrium dynamical systems with application to biology},
  journal = {Advances in Physics},
  volume  = {64},
  number  = {1},
  pages   = {1-137},
  year    = {2015},
  doi     = {https://doi.org/10.1080/00018732.2015.1037068}
}

@Article{times,
  author  = {Andrew~H. Knoll and Martin~A. Nowak},
  title   = {The timetable of evolution},
  journal = {Science Advances},
  volume  = {3 (5)},
  pages   = {e1603076},
  year    = {2017},
  doi     = {https://doi.org/10.1126/sciadv.1603076},
}

@Article{Akaike,
  author  = {Joseph~E. Cavanaugh and Andrew~A. Neath},
  title = {The {A}kaike information criterion: {B}ackground, derivation, properties, application, interpretation, and refinements},
  journal = {WIREs Comput Stat.},
  volume = {11},
  number = {3},
  pages = {e1460},
  year = {2019},
  doi     = {https://doi.org/https://doi.org/10.1002/wics.1460},
}

@Article{ergodic,
  author  = {Calvin~C. Moore},
  title   = {Ergodic theorem, ergodic theory, and statistical mechanics},
  journal = {Proceedings of the National Academy of Sciences},
  volume = {112},
  number = {7},
  pages = {1907-1911},
  year = {2015},
  doi = {https://doi.org/10.1073/pnas.1421798112}
}

@Article{Semiclass,
  author = {M V Berry},
  title = {Evolution of semiclassical quantum states in phase space},
  journal = {Journal of Physics A: Mathematical and General},
  volume = {12},
  number = {5},
  pages = {625-642},
  year  = {1979},
  doi = {https://doi.org/10.1088/0305-4470/12/5/012}
}

@Book{gaussian,
  author  = {Ariel Caticha},
  title   = {Entropic inference and the foundations of physics},
  publisher = {Brazilian Chapter of the International Society for Bayesian Analysis-ISBrA, Sao Paulo, Brazil},
  howpublished = {\url{http://dl.icdst.org/pdfs/files1/77964f05542451c01e8e420e975dd664.pdf}},
  year    = {2012}
}

\section*{ Acknowledgments}
A.G. acknowledges the Cuban Program for Basic Sciences, the Office of External Activities of the Abdus Salam Centre for Theoretical Physics, and the University of Electronic Science and Technology of China for support. The research is carried on under a  project of the Platform for Bio-informatics of BioCubaFarma, Cuba. The data for the present analysis come from the TCGA Research Network: \url{https://www.cancer.gov/tcga} \cite{tcga}. Authors are grateful to Gabriel Gil for comments and a critical reading of the manuscript.

\section*{Author's contributions}
A.G. conceived and coordinated the work. F.Q and A.G. processed the experimental data. 
F.Q. and D.A.L. contributed to the GitHub repository. M.L.B. and P.V.S. introduced the information theory concepts. All authors analyzed and 
interpreted the results, contributed to the manuscript and approved the final version.

\section*{Competing interests}
  The authors declare that they have no competing interests.

\section*{Availability of data and materials}

	The information about the data we used, the procedures and results are integrated in a public repository that is part of the project "Processing and Analyzing Mutations and Gene Expression Data in Different Systems": \url{https://github.com/DarioALeonValido/evolp}.
	
	To process the data set we include a script in \path{../evolp/Entropy_Tumors/}. The script reads the TCGA data replicated in the folder \path{../databases_external/TCGA/} and the data coming from the PCA analysis located in path \path{../databases_generated/TCGA_pca/}. 




\section*{Supplementary material}

# Supplementary information for: "Estimating the number of available states for normal and tumor tissues in gene expression space"

Augusto Gonzalez[1,2], Frank Quintela[3,2], Dario A. Leon[4,2], Maria Luisa Bringas Vega[1,5], Pedro Valdes Sosa[1,5]

[1]University of Electronic Science and Technology, 610051, Chengdu (People Republic of China)
[2]Institute of Cybernetics, Mathematics and Physics, 10400, Havana (Cuba)
[3]University of Modena & Reggio Emilia, 41125, Modena, (Italy)
[4]S3 Centre, Istituto Nanoscienze, CNR, 41125, Modena (Italy)
[5]Cuban Neurosciences Center, 11600, Havana, (Cuba)

November 13, 2021

## Abstract

We provide supplementary details on the way the calculations were performed.

## The TCGA terminology

The exact meaning of TCGA abbreviations for cancer types is provided in Supplementary Table S1.

## The effective dimension of the region spanned by the samples

In Supplementary Table S1, we included the fraction of total variance captured by the first 20 principal components in each tumor. These numbers are above 80 % in all cases. This suggests that the effective dimension of the region spanned by the samples is around 20.

This idea is reinforced by Supplementary Fig. S1 upper panel, where the total variance for BRCA is plotted as a function of the number of principal components. In addition, we may use arguments from information theory [1] in support of the fact that the effective dimension is near 20.

As mentioned in the Methods section of the main manuscript, in the PC analysis deviations and variances are measured with respect to $\hat{e} = 0$. Thus, in each PC axis our model for the data uses a Gaussian with zero mean and standard deviation $\sigma$, that is, a single parameter for each dimension. The Akaike Information Criterion (AIC) is a compromise between twice the number of used parameters and the logarithm of the Maximum Likelihood function, $L$. The number of parameters for each dimension is $2 - \ln L(\sigma)$, where $\ln L(\sigma) = \sum_i \{-0.5 \ln(2\pi) - \ln(\sigma) - 0.5(x_i/\sigma)^2\}$ and the sum runs over the samples.

Let us assume that we compute the first 200 components of PCA. The AIC for our model is:

$$AIC_1 = \sum_d \{2 - \ln L(\sigma_d)\},$$

where $d$ runs from 1 to 200, and $\sigma_d$ is the standard deviation of the sample positions along the $d$ axis.

Let us introduce a second model, with an effective dimension $d_e$. It means that for the dimensions $d > d_e$ we will use a unique parameter, $\sigma_e$. Our second model has only $d_e + 1$ parameters, and its AIC is:

$$AIC_2 = \sum_{d \le d_e} \{2 - \ln L(\sigma_d)\} + 2 - \sum_{d > d_e} \ln L(\sigma_e).$$

Of course, it is always more precise and informative to use an independent $\sigma_d$ for each $d$, that is $AIC_1 < AIC_2$, and the first model is preferred. However, as it is illustrated in Supplementary Fig. S1 bottom panel for Breast Cancer (BRCA), there is a fast saturation of $AIC_2$ as a function of $d_e$. For $d_e > 15$, $AIC_2$ is greater than 0.96 $AIC_1$. There is a minimal gain from the point of view of information theory in using an effective dimension greater than 20.

In the paper we, for the first time, estimate the volume of the region spanned by the samples in gene expression space using as effective dimension $d_e = 20$. We compare the normal and tumor regions for a given tissue, and compare also these magnitudes for different tissues.

## Dependence of $S$ and $I$ on the number of samples used in their computation

As already mentioned in the main text, the choice of $\rho$ in Eq. (6) as a multivariate gaussian has many advantages, among them the fact that, according to the Central Limit Theorem, it improves with the increase of the number of samples.

We show in Supplementary Fig. S2 this dependence for the entropy of normal and tumor regions in breast cancer (BRCA). From now on, computations use $d_e = 20$. Each point in the figure corresponds to an average over 1000 sets of $n$ samples, randomly selected among the original samples. It is apparent that around 250 samples are needed to obtain converged results. In the data, tumor samples satisfy this requirement, that is the error in the computation of their entropy is practically zero. But normal samples are usually well below these numbers. The error in the computation of entropy differences comes mainly from the entropy of the normal region.

The center panel shows that the entropy difference exhibits a smooth dependence on $n$. In the paper, a crude estimation of the error for S(normal), and thus of $\Delta S$, is obtained from the standard deviation of 1000 realizations of S(tumor), computed with a number of samples equal to the normal ones.

The same procedure is used for the overlap, $I$. This is a more subtle magnitude. Nevertheless, the convergence of $-\ln(I)$ as a function of the number of samples is also smooth, as shown in the bottom panel of the figure. A crude estimation of the error is obtained from the standard deviation in 1000 random sets of $n$ samples for $\rho_{tumor}$, where $n$ is the number of normal samples.

## References

[1] Joseph E. Cavanaugh and Andrew A. Neath. "The Akaike information criterion: Background, derivation, properties, application, interpretation, and refinements". In: *WIREs Comput Stat.* 11.3 (2019), e1460. DOI: https://doi.org/https://doi.org/10.1002/wics.1460.

| Abbreviation | Cancer type | Total variance |
|---|---|---|
| BRCA | Breast invasive carcinoma | 88.86% |
| COAD | Colon adenocarcinoma | 89.04% |
| HNSC | Head and and neck squamous cell carcinoma | 86.40% |
| KIRC | Kidney clear cell carcinoma | 83.51% |
| KIRP | Kidney papillary cell carcinoma | 87.02% |
| LIHC | Liver hepatocellular carcinoma | 85.46% |
| LUAD | Lung adenocarcinoma | 87.84% |
| LUSC | Lung squamous cell carcinoma | 90.09% |
| PRAD | Prostate adenocarcinoma | 84.92% |
| STAD | Stomach adenocarcinoma | 84.93% |
| THCA | Thyroid carcinoma | 89.48% |
| UCEC | Uterine corpus endometrial carcinoma | 88.39% |

Table S1: TCGA abbreviations for the set of studied cancer types. The total variance captured by the first 20 principal components is also included in the table.

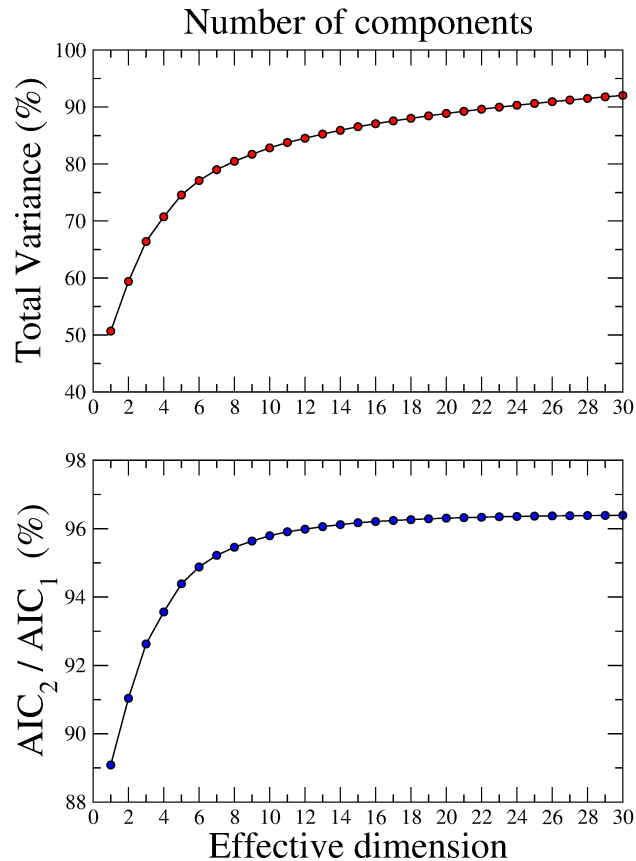

Figure S1: Upper panel: Total variance in breast cancer (BRCA) as a function of the number of components in a PCA calculation. Lower panel: Saturation of the AIC as a function of the effective dimension (see explanation in the main text).

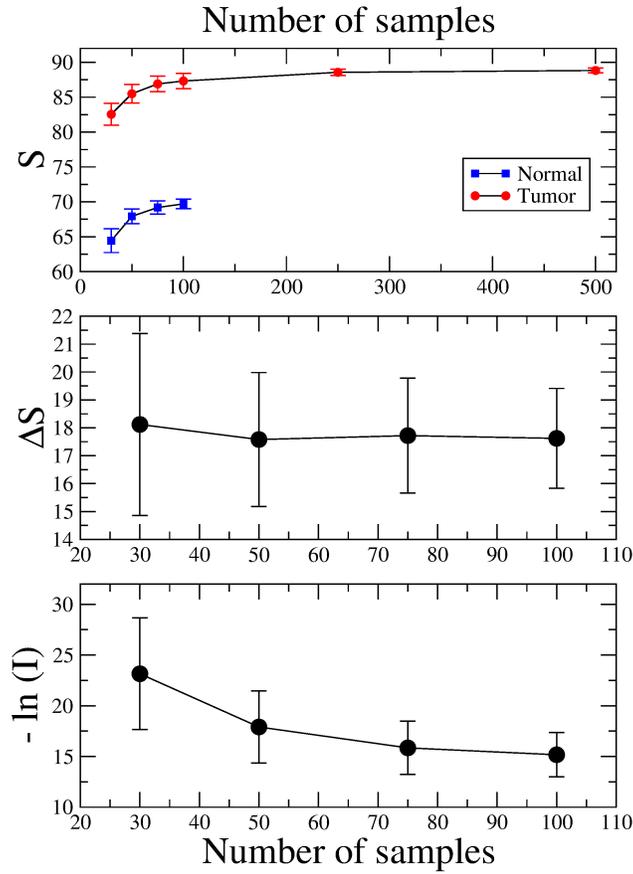

Figure S2: Upper panel: The entropy of tumor and normal samples in breast cancer (BRCA) as a function of the number of samples used in their computation. Each point represent the average over 1000 sets of randomly selected $n$ samples. Converged result are obtained with $n \geq 250$ samples. Center panel: The entropy difference exhibits a smooth dependence on $n$. Lower panel: The dependence of $-\ln(I)$ on $n$.



\end{document}